\documentclass[twocolumn]{aastex6}

\usepackage[english]{babel}

\usepackage[T1]{fontenc}
\usepackage{ae,aecompl}

\usepackage{fancyhdr}
\usepackage{amsfonts}
\usepackage{amsmath}
\usepackage{amssymb}
\usepackage{layout}
\usepackage{graphicx}
\usepackage{multirow}
\usepackage{color}
\usepackage{multirow}

\usepackage{times}
\usepackage{natbib}
\def\be{\begin{equation}}
\def\ee{\end{equation}}

\newif\ifAMStwofonts
\AMStwofontstrue

\graphicspath{{./fig/}}

\shorttitle{model selection}
\shortauthors{Rezaei et al.}

\begin{document}

\title{Comparison between different methods of model selection in cosmology}
\author{ Mehdi Rezaei \altaffilmark{1}, \altaffilmark{2}}
\author{ Mohammad Malekjani \altaffilmark{1}}

\affil{\altaffilmark{1} Department of Physics, Bu-Ali Sina University, Hamedan 65178, Iran }

\affil{\altaffilmark{2} Iran meteorological organization, Hamedan Research Center for Applied Meteorology, Hamedan, Iran }


\begin{abstract}
There are several methods for model selection in cosmology which have at least two major goals, that of finding the correct model or predicting well. In this work we discuss through a study of well-known model selection methods like Akaike information criterion (AIC), Bayesian information criterion (BIC), deviance information criterion (DIC) and Bayesian evidence, how these different goals are pursued in each paradigm. We also apply another method for model selection which less seen in cosmological literature, the Cross-validation method. Using these methods we will compare two different scenarios in cosmology, $\Lambda$CDM model and dynamical dark energy. We show that each of the methods tends to different results in model selection. While BIC and Bayesian evidence overrule the dynamical dark energy scenarios with 2 or 3 extra degree of freedom, the DIC and cross-validation method prefer these dynamical models to $\Lambda$CDM model. Assuming the numerical results of different analysis and combining cosmological and statistical aspects of the subject, we propose cross-validation as an interesting method for model selection in cosmology that can lead to different results in comparison with usual methods of model selection. 
	
\end{abstract}
\maketitle

\section{Introduction}\label{sect:int}

The evidence of the accelerated expansion of current Universe firstly comes from extensive surveys of
high-redshift type Ia supernovae (SnIa) in the late 1990s \citep{Riess1998,Perlmutter1999}. Recent advances in observational cosmology also confirms this acceleration phase. In order to explain this accelerating phase, cosmologists have proposed many different cosmological models. These models can be divided into two main groups based on theoretical assumptions on the cause of the accelerated expansion.
In the first group, the accelerated expansion can be explained by introducing an unknown form of component with negative pressure, dubbed dark energy (DE),  while in the second group of models the mentioned expansion is justified by some modification of the standard theory of gravity on extragalactic scales \citep[see][]{Carroll:2003wy,Kowalski2008,Nojiri:2010wj}.\\
In the framework of general relativity (GR), the cosmological constant $\Lambda$ is the simplest and most likely candidate for DE and occupies about $70 \%$ of the energy budget of the universe \citep{Bennett:2003bz,Peiris:2003ff,Spergel:2003cb}.
One can easily explain the recent accelerated expansion phase of the universe by assuming the cosmological constant ($\Lambda$) and cold dark matter (CDM). Besides the observational success of the $\Lambda$CDM model, it suffers from some theoretical problems such as the fine-tuning problem and the coincidence problem \citep{Weinberg1989,Padmanabhan2003,Copeland2006}.\\ Furthermore, the $\Lambda$CDM model plagued with some observational tensions in estimation of some key cosmological parameters. In particular as the first tension, the Lyman-$\alpha$ forest measurement of the Baryon Acoustic Oscillations from \citep{Delubac:2014aqe}, suggests a smaller value of matter density parameter in comparison with that of obtained by CMB data. Furthermore, the other tension concerns the discrepancy between large scale structure data from \citep{Macaulay:2013swa}, and too large value of $\sigma_8$ predicted by the $\Lambda$CDM. As the other tension, there is a statistically significant disagreement between the $H_0$ value obtained by the classical distance ladder and that measured by the Planck CMB data \citep{Freedman:2017yms}.\\
 Quantitatively speaking, the $\Lambda$CDM cosmology deduced from Planck CMB data predict $H_0 = 67.4\pm0.5$ km/s/Mpc \citep{Aghanim:2018eyx}, while from the Cepheid-calibrated SnIa \citep{Riess:2019cxk} we have $H_0=74.03\pm 1.42$ km/s/Mpc. Some investigations using model independent cosmographic approaches in the literature show that this big value of tensions which observed between the results of $\Lambda$CDM and those obtained in model independent approach, support this claim that we should investigate for finding good alternatives to $\Lambda$CDM \citep[see also][]{Rezaei:2017hon,Yang:2019vgk,Khadka:2019njj,Lusso:2019akb,Benetti:2019gmo}.\\
 Exiting from the $\Lambda$CDM model proposed different kinds of DE models, from simplest generalization of the $\Lambda$CDM, the so-called $w$CDM model, which assumes DE as a perfect fluid with a constant state parameter $w$ differs from $-1$ to other different dynamical DE scenarios in literature \citep[see also][]{Veneziano1979,Erickson:2001bq,Thomas2002,Caldwell2002,Padmanabhan2002,Gasperini2002,Elizalde:2004mq,Gomez-Valent:2014fda}. However, in order to solve some of tensions in $\Lambda$ cosmology, we need some thing more than late time deviations from $\Lambda$CDM. For instance, to solve the Hubble tension we require modifications prior to recombination which lower the sound horizon in order to match BAO and uncalibrated SnIa observations, as discussed in \citep{Knox}.\\
Although, roughly all of these models can justify the historical evolution of the universe, especially the current accelerated expansion, but just some of them can solve or alleviate the problems of $\Lambda$CDM. To compare these models and find the best of them, one can do a comparison between them with different observational data and examine those compatibility with observations. In the literature, it is common to use different approaches to select the best model among different ones. Some of these approaches which more be used by cosmologists are the least squares method,  likelihood-ratio test,  Akaike information criterion (AIC)\citep{Akaike:1974}, Bayesian information criterion (BIC)\citep{Schwarz:1974}, deviance information criterion (DIC)\citep{dicref} and statistical Bayesian evidence \citep[see also][]{Malekjani:2016edh,Rezaei:2017yyj,Malekjani:2018qcz,Lusso:2019akb,Rezaei:2019roe,Lin:2019htv,Rezaei:2019xwo,Rezaei:2020mrj,Rezaei:2020lfy}. Also the Bayesian model comparison approach as applied to cosmological models has
been strongly criticized by \cite{Cousins:2008gf,Cousins:2013hry}. Doing a comparison between cosmological constant and the dynamical DE shows that the Bayesian Evidence will only be of interest in model selection, if the models and priors are physically well motivated \citep{Efstathiou:2008ed}. However, Bayesian evidence remains the standard for model comparison in the field, much more than other methods. Moreover, there are some new methods to overcome the limits of Bayesian evidence, for instance the open likelihoods method which discussed in \citep{Gariazzo}.
 Thus it can be useful if one can find more proper approaches for model selection. In the field of statistics, there is an other conventional approach for model selection, dubbed cross-validation. Different methods of this approach have been investigated in \citep{dabbs2016comparison} and those results have been compared with AIC and BIC. It has been shown that the cross-validation performs more accurate model selection, and avoids over-fitting better than any of the other model selection methods considered in \citep{dabbs2016comparison}.\\
 In this work we want to use cross-validation method for model selection and compare it with other common approaches in cosmology. In this way we will assume two different parametrizations to introduce the Equation of State (EoS) parameter of DE and compare those results with concordance $\Lambda$CDM. These parametrizations can help us in order to study the main effects of dynamical DE in comparison with cosmological constant. More details about these parametrizations can be seen in section \ref{sect:md}. Previously in \citep{Rezaei:2017yyj}, these DE scenarios have been compared in the light of different observational data combinations using both of AIC and BIC criteria. The aim of this paper is to compare these models by cross validation method. In Sect.\ref{sect:ms}, we briefly introduce the cross validation and other different approaches for model selection which commonly used in cosmology and some details of the procedure which we have applied in our analysis. In Sect.\ref{sect:md}, firstly we describe the basic equations of cosmological models under study and secondly we present our data samples. We present our results and discussions in Sect.\ref{sect:discus}. Sect.\ref{sect:conc} briefly gives the conclusions.

\section{How to select the best model?}\label{sect:ms}

To compare different cosmological models in the light of the observational data we need procedures which can numerically determine the goodness of fit and also the best value of free parameters. One of the simplest methods which have commonly been applied in cosmology is the least squares method. It is a statistical procedure to find the best fit for a set of data points by minimizing the sum of the offsets or residuals of data points from the model values \citep{Aldrich}. Using this method we can find the best fit parameters of a model and its goodness of fit. Although increasing the number of free parameters can improve the goodness of fit of the models to data, but it also enhances the complexity of the model.\\
In comparing different models, the least squares method can not remove the effects of additional free parameters. Therefore in comparing various models with different number of free parameters we should choose another approach. To solve this problem, it is proposed some criteria, based on Occam's razor which says " among competing hypotheses, the one with the fewest assumptions should be selected". Given a choice of competing theories, Occam's razor is the principle that directs us to pick the simplest one as the most likely to be correct\citep{ralph2015combinatorics}. Based on this principle Akaike proposed a criteria the so-called Akaike Information Criteria (AIC)\citep{Akaike:1974}.
Historically AIC is the first penalized maximum likelihood rule. If one reads Akaike's early papers and resolves the ambiguities in the light of the further pioneering theoretical from \citep{Shibata1984}, it becomes clear that the objective is to choose a theoretical model so that one can predict optimally the data on the dependent variables in an exact new replicate of the design for the given data. In fact, the AIC is maximized over models to get the best predictive model. Furthermore  Shao's theorem \citep{shao1997asymptotic} shows that AIC can help us to identify the most useful model from the point of view of prediction \citep{dutta2015model}.\\
Bayes Information Criterion (BIC)\citep{Schwarz:1974} as the second penalized maximum likelihood rule, provides a convenient approximation,which may be interpreted as a penalized maximum likelihood corresponding to a model. BIC is very advantageous and can be used to choose the correct model with probability tending to one \citep{dutta2015model}. It has been known for quite some time that Bayesian estimation of free parameters or predicting the future observations is completely robust with respect to the choice of prior while model selection based on Bayes factors is not robust. On the other hand, BIC as an approximation gives good results if in the set of the models under consideration is one favored model. The problems appear when we have two favored models with nearly the same probabilities\citep{Kurek:2007tb}.
Moreover, in \citep{Arevalo:2016epc,Rezaei:2019xwo}, authors noted some tensions between the results which they have obtained from AIC and BIC.\\ 
Deviance Information Criteria (DIC) as another method of model selection, has been developed by Bayesian authors in \citep{dicref}, in an effort to generalize AIC, using the Kullback-Liebler divergence instead of squared error loss. To some extent this has been done earlier also. However, DIC also tries to find a form of penalty that can take care of hierarchical models as well as latent parameter models. It seems DIC is really suitable for prediction but it has also been used for choosing the correct model. Also, there have been doubts about the penalty. Interested readers should read both the discussion and the reply of the authors \citep{dutta2015model}.
As we mention before, the Bayesian model comparison method, has been criticized in some references in which, pointed out that often insufficient attention is given to the selection of models and of priors, and that this might lead to posterior model probabilities which are largely a function of one's unjustified assumptions. This draws attention to the difficult question of how to choose priors on phenomenological parameters, for which theoretical reasoning offers poor or no guidance \citep{Trotta:2017wnx}. Bayesian model selection also returns the best model among the set of different models being compared, even though that model might be a poor explanation for the given data. So we should look for a  principled way of constructing an absolute scale for model performance in a Bayesian context\citep{March:2010ex,Trotta:2017wnx}.\\
Thus in this work we do our analysis using another method of model selection, the cross-validation. Cross-validation is a popular non-parametric method for evaluating the accuracy of a predictive model\citep{wager2019crossvalidation}.
The results of \citep{wager2019crossvalidation} shows that cross-validation is asymptotically uninformative about the expected test error of any given predictive rule, but allows for asymptotically consistent model selection. This phenomenon appears because the leading-order error term of cross-validation doesn't depend on the model being evaluated, and so canceled out when one compares two models. Furthermore, cross-validation help us to prevent two significant problems in model selection: Over-fitting vs. Under-fitting\citep{ghojogh2019theory}.\\
Here we use K-fold cross-validation method to have a comparison among different DE scenarios. Moreover, we can compare the results of cross-validation with those we obtain from AIC, BIC, DIC and the Bayesian evidence methods. In k-fold cross-validation method the data are randomly partitioned into $k$ equal-size subsets. In turn each of the $k$ subsets is retained as the validation set, while the remaining $k-1$ folds work as the training set, and the average prediction error of each candidate model is obtained \citep{arlot2010}. In the literature, there are conflicting recommendations on the data splitting ratio for cross-validation  \citep{arlot2010}.
As the data can be divided in many ways into $k$ groups, it introduces additional variance in the estimates. This variance can be reduced by repeating k-fold cross-validation several times with different permutations in the data division. When $k$ is increasing, we observe an decrement in the training error. While as $k$ increases, the cross-validation error first decreases and then starts increasing. Therefore, we try to find the optimum value of $k$ which minimizes the cross-validation error.
In order to find this optimum value, we repeated our analysis using different values of $k$ between $5<k<15$. We find that for the values of $k$ between 8 to 11, the cross-validation error reaches its minimum. This result is in agreement with those obtained in literature\citep[see also][]{arlot2010,Vehtari,kohavi,zhang2015}. Kohavi focused only on accuracy estimation in all the numerical work in \citep{kohavi}. His work
has been very well-known and the recommendation there that the best method to use for model selection is 10-fold Cross Validation has been followed by many in computer science, statistics and other fields\citep{zhang2015}.

Upon our results and in the line of mentioned literature, we use 10-fold cross-validation for selecting the best model. In this way we have divided our data points in to ten sub-samples. In order to split our data points we sort all of data points upon redshift, select the first to 10th data points from the list and put each of them in a different sub-samples. Then we repeat this procedure for 11th to 20th data point of the list and follow this procedure to the end of the list of data points.
After preparing sub-samples in each step of the analysis, we leave one of our sub-samples out (test sample) and using other ones (training sample) to put constraints on the free parameters of the model. Now, based on the best fit values, we compare theoretical values predicted by model with the test sample and find its $\chi^2_t$ value. This process repeat again until all of the sub-samples will be used as a test sample.\\
Finally, we calculate the sum of $\chi^2_t$ values to obtain $\chi^2_{tot}$. Doing this procedure for each of the models, we will have a $\chi^2_{tot}$ value which shows the goodness of fit for that model. The model with the lowest cross-validation score ($\chi^2_{tot}$ in our analysis) will perform best on the testing data and will achieve a balance between under-fitting and over-fitting. To have a comparison between cross-validation and commonly used criteria, AIC and BIC, we perform the Markov chain Monte Carlo (MCMC) analysis and upon the results of this analysis we will compute AIC and BIC values for each of the models under study. For more details concerning the MCMC analysis we refer the reader to \cite{Mehrabi:2015hva} \citep[see also][]{Malekjani:2016edh,Rezaei:2017yyj,Rezaei:2019roe}. Moreover, we compute the other criteria, DIC and the well known Bayesian evidence \citep[for more details see][]{Rezaei:2020mrj} for different cosmological scenarios. In the next section we introduce the cosmological models that we want to investigate and also represent the observational data sets used in our analysis. 

\section{Models and data sets}\label{sect:md}
One of the simplest ways for studying the EoS parameter of dynamical DE models is via a parameterization. It is easy to find many different EoS parameterizations in literature. The earliest parameterization is the Taylor expansion of EoS with respect to $z$ up to first order \citep{Maor:2000jy,Riess:2004nr}.  This simple parametrization also have be generalized by considering the second order approximation in Taylor series \citep{Bassett:2007aw}.  Although these two parameterizations can introduce the dynamical properties of DE in a very simple way, but they diverge at higher redshifts. Thus, investigations have been continued for finding more suitable parameterizations. The suitable parameterizations should can provide the dynamical properties of DE EoS and also prevent any divergence at high redshifts.
In order to achieve this goal, some purely phenomenological parameterizations have been proposed in literature \citep{Efstathiou:1999tm,Jassal:2004ej,Barboza:2009ks}. Here we consider two well known parameterizations namely CPL and PADE parameterizations as the alternatives of cosmological constant. 
For a good behavior function $f(x)$, the PADE approximation of order $(i,j)$ has the following form \citep{pade1892,baker96,Adachi:2011vu}:

\begin{eqnarray}\label{padeO}
f(x)=\frac{a_0+a_1x+a_2x^2+...+a_jx^j}{b_0+b_1x+b_2x^2+...+b_ix^i}\;,
\end{eqnarray}
where all the coefficients are constants. In this study, as the PADE parameterization, we focus on the expansion of the EoS parameter $w_{\rm d}$ with respect to $(1-a)$ up to order $(1,1)$ upon Eq.(\ref{padeO}) as:
\begin{eqnarray}\label{pade1}
w_{\rm d}(a)=\frac{w_0+w_{1}(1-a)}{1+w_{2}(1-a)}\;.
\end{eqnarray}

To study the background evolution of the universe in PADE cosmology, we assume an isotropic and homogeneous spatially flat FRW cosmologies. In this cosmology which driven by non-relativistic matter and DE, the first Friedmann equation reads 
\begin{eqnarray}\label{frid1}
H^2=\frac{8\pi G}{3}(\rho_{\rm m}+\rho_{\rm d})\;,
\end{eqnarray}
where $H\equiv {\dot a}/a$ and $\rho_{\rm m}$ and $\rho_{\rm d}$ are the energy densities of dark matter and DE, respectively. In the absence of interaction among the cosmic fluid components, we will have the following differential equations for different energy densities:
 \begin{eqnarray}\label{continuity}
 &&\dot{\rho_{\rm m}}+3H\rho_{\rm m}=0\;,\label{matter}\\
&&\dot{\rho_{\rm d}}+3H(1+w_{\rm d})\rho_{\rm d}=0\;\label{de},
 \end{eqnarray}
where the over-dot indicates a derivative with respect to cosmic time $t$. 
Inserting Eqs . (\ref{pade1}) into equation (\ref{de}), we can find the DE density of PADE parameterizations 
\citep[see also][]{Rezaei:2017yyj,Rezaei:2019hvb}

\begin{eqnarray}
\rho_{\rm d}^{(\rm P)}=\rho_{\rm d0} a^{-3(\frac{1+w_0 + w_1+ w_2}{1+w_2})} [1+w_2 (1-a)]^{-3(\frac{w_1- w_0 w_2}{w_2 (1+w_2)})}\;,\label{rho-pade1}
\end{eqnarray}
Combining Eq.(\ref{rho-pade1}) with Eq.(\ref{frid1}) we obtain the dimensionless Hubble parameter $E=H/H_0$ as follow\citep[see also][]{Rezaei:2017yyj}: 

\begin{eqnarray}
&&E^{2}_{\rm (P)}=\Omega_{\rm m0} a^{-3} + \Omega_{\rm d0}a^{-3(\frac{1+w_0+ w_1 + w_2}{1+w_2})}  \times \nonumber
\\
&&(1+w_2 - a w_2)^{-3(\frac{w_1 -w_0 w_2}{w_2(1+w_2)})} \;,\label{Epade1}
\end{eqnarray}
where $\Omega_{\rm m0}$ is the matter density parameter and thus we have $\Omega_{\rm d0}=1-\Omega_{\rm m0}$ as DE parameter.
As the second DE parametrization in this work, we investigate the CPL parameterization. One can easily find that setting $w_2=0$, Eq. (\ref{pade1}) reduces to CPL parameterization. We note that unlike CPL parameterization, in PADE cosmology, the EoS parameter with $w_2\neq 0$ avoids the divergence at far future ($a\to\infty$).
Following the above lines, for CPL parameterization, it is easy to find 
\begin{eqnarray}\label{rho-cpl}
\rho^{\rm (C)}_{\rm d}=\rho_{\rm d0}a^{-3(1+w_0+w_1)}\exp[{-3 w_1 (1-a)}]\;,
\end{eqnarray}
and 
\begin{eqnarray}
E^{2}_{\rm (C)}=\Omega_{\rm m0} a^{-3} + \Omega_{\rm d0} a^{-3(1+w_0+ w_1)}\exp[{-3 w_1 (1-a)}]\;.\label{Ecpl}
\end{eqnarray} 
Now, in order to study the performance of standard $\Lambda$CDM and the above cosmological parameterizations against observations, we assume two different sets of latest data samples:
\begin{itemize}
\item Pantheon: This sample includes the apparent magnitude of 1048 type Ia supernovae in the range of $0.01 < z < 2.26$  \citep{Scolnic:2017caz}. Using this sample we can constrain cosmological parameters through the comparison  of their apparent luminosities.
\item Hubble data: Latest measurements from cosmic chronometers for Hubble parameter,$H(z)$, from \citep{Farooq:2016zwm} is the second sample which we have used in our analysis.
\item
Baryon acoustic oscillations (BAO): In order to constrain late-time new physics we have used 6 data points of available BAO measurements from 6df Galaxy,MGS, BOSS DR12, SDSS DR12, BOSS DR14 and DES collaborations \citep{bao55,bao37, bao56,bao57,bao58,bao39}.
\end{itemize}

\section{Results and discussions}\label{sect:discus}
In order to obtain the value of AIC and BIC values for DE models under study, we perform statistical analysis using  mentioned data samples in MCMC algorithm. In particular the total chi-square $\chi^2_{\rm t}$ has the following form: 

\begin{eqnarray}\label{eq:like-tot_chi}
\chi^2_{\rm t}({\bf p})=\chi^2_{\rm Pantheon}+\chi^2_{\rm Hubble}+\chi^2_{\rm BAO}\;.
\end{eqnarray}
where ${\bf p}$ is the vector including the free parameters of the model under study. The relevant vectors for different cosmological models in this work are:

\begin{itemize}
\item PADE: ${\bf p}=\{\Omega_{\rm M0}, h,w_0,w_1,w_2\}$
\item CPL: ${\bf p}=\{\Omega_{\rm M0}, h,w_0,w_1\}$
\item $\Lambda$CDM: ${\bf p}=\{\Omega_{\rm M0}, h\}$
\end{itemize}
Computing the minimum value of $\chi^2_{\rm t}$ ($\chi^2_{\rm min}$), we can obtain the value of AIC and BIC for each of the model under study:
\begin{eqnarray}
{\rm AIC} = \chi^2_{\rm min}+2K\;,\\
{\rm BIC} = \chi^2_{\rm min}+K\ln N\;,
\end{eqnarray}

where $K$ and $N$ are the number of free parameters and the total number of data points respectively.
The DIC criterion employs both Bayesian statistics and information theory concepts \citep{dicref},
and it is expressed as \citep{ref0701}
\begin{eqnarray}
{\rm DIC} = D(\bar{{\bf p}})+2C_{\rm B}\;.
\end{eqnarray}

where the quantity $C_{\rm B}=\bar{{D({\bf p})}}-D(\bar{{\bf p}})$ is the Bayesian complexity and over-lines imply the standard  mean value. Moreover, $D({\bf p})$ is the Bayesian deviation, which can be expressed as$D({\bf p})=\chi^2_{\rm t}({\bf p})$ in the case of exponential class of distributions \citep[see more details in]{ref1604,ref0701}. It is closely related to the $K$, number of effective degrees of freedom, which is actually the number of parameters that affect the fitting. In a less strict manner, it could be considered as a measure of the spread of the likelihood \citep{dicmain}.

As a different model selection method we compute the Bayesian evidence in different scenarios which we consider in this paper. Considering $\Theta$ as the free parameters of the model $M$ and $D$ as data set, Bayesian evidence $\epsilon$ is given by:

\begin{eqnarray}
\epsilon = p(D \mid M)=\int p(\Theta \mid M)p(D \mid \Theta,M)d\Theta\;.
\end{eqnarray}
Although this might has an analytic solution for a low dimensional cases, for a high denominational problem it is intractable analytically and we have to use numerical methods to evaluate the integral value. Here, we use the sequential Monte Carlo (SMC) algorithm to sample the posterior. The evidence is a crucial quantity for model selection in Bayesian framework and in comparison between two models.
In this paper, we use the Jeffreys’ scale \citep{ref47} to measure the significant difference between two different models. Considering two models $M_1$ and $M_2$ the Jeffreys scale with respect of $\Delta \ln \epsilon = \ln \epsilon_{ M_1} -\ln \epsilon_{ M_2}$ is as following\citep{ref48}:

\begin{itemize}
\item for $\Delta \ln \epsilon < 1.1$ there is a weak evidence against $M_2$.
\item for $1.1<\Delta \ln \epsilon < 3$ there is definite evidence against $M_2$.
\item for $\Delta \ln \epsilon > 3$ there is strong evidence against $M_2$ 
\end{itemize}

We report the numerical results of our analysis for different models under consideration in Tab.\ref{tab:best}.

\begin{table*}
	\centering
	\caption{The statistical results for different cosmological models studied in this work obtained using MCMC analysis. Note that we calculate the values of $\Delta$AIC, $\Delta$BIC and $\Delta$DIC related to the $\Lambda$CDM. Moreover, in the last column we have $\Delta \ln \epsilon= \ln \epsilon_{\Lambda CDM}-\ln \epsilon_{Model}$ as the result of Bayesian evidence.}
	\begin{tabular}{c  c  c c c c c c c c c}
		\hline \hline
		Model  & $K$ & $N$ & $\chi^2_{min}$& AIC & $\Delta$AIC & BIC & $\Delta$BIC & DIC & $\Delta$DIC & $\Delta \ln \epsilon$  \\
		\hline
		PADE & $5$ & $1092$ & $1056.6$ & $1066.6$ & $+1.2$ & $1091.6$ & $+16.2$& $1065.9$ & $0.0$ & $2.2$ \\
		\hline
		CPL & $4$ & $1092$ & $1057.6$ & $1065.6$ & $+0.2$ & $1085.6$& $+10.2$ & $1068.0$ & $+2.1$ & $2.7$  \\
		\hline
		$\Lambda$CDM & $2$ & $1092$ & $1061.4$ & $1065.4$ & $0.0$  & $1075.4$ & $0.0$ & $1078.3$ & $+12.4$ & $0.0$  \\
		\hline \hline
	\end{tabular}\label{tab:best}
\end{table*}

As one can see in the Tab.\ref{tab:best}, value of $\chi^2_{min}$ which we have obtained for each of the models indicates that the related model can be a good model for fitting the observational data. But as we mentioned before, because of different number of free parameters, the values of $\chi^2_{min}$ is not suitable for comparing these models. Therefore, in the next step we use AIC values for model comparison. As we can see in numerical results,$\Lambda$CDM with $AIC=1065.4$ has the least value of AIC and upon this criteria we can say $\Lambda$CDM is the best model and on the other hand, PADE is the worst one. But, how we can determine the distance between the best model and worst one from AIC point of view. The value of $\Delta$AIC is the parameter that determine this distance. Although $\Lambda$CDM with the minimum value of AIC is the best model, but $\Delta AIC < 2$ for two other models indicate that there are \textit{Significant support} to CPL and PADE  models. Thus we can conclude from AIC results that $\Lambda$CDM is the best model and also we observe significant support to other models under study. For more details concerning $\Delta AIC$ and the level of support to a model upon it we refer the reader to \citep{Kass:1995loi,Rezaei:2019roe}.    
In the BIC column we observe that $\Lambda$CDM has the minimum value. In other words, from BIC point of view the $\Lambda$CDM is the best model and PADE is the worst one. This result was expected because of existence of more extra free parameters in PADE parameterization. The results which we find from $\Delta$BIC values are completely different from those we find from $\Delta$AIC. In this case we have $\Delta BIC_{PADE}=16.2$ and thus \textit{Very strong evidence against} PADE parameterization. For CPL parameterization we have $\Delta BIC_{CPL}=10.2$ and equivalently \textit{Very strong evidence against} this cosmology. Reader can find more details concerning $\Delta BIC$ and the strength of evidence against each candidate model in Refs.\citep{Kass:1995loi,Rezaei:2019roe}. Here we observe a significant conflict between the conclusions we obtained from AIC and BIC criteria.\\
Assuming the values of $\Delta$DIC for different models we obtain different results compared with those we obtained from AIC and BIC. From the value of $\Delta DIC= 12.4$ we observe very strong evidence against $\Lambda$CDM, while PADE with minimum value of DIC is the best model.
From the Bayesian evidence point of view, $\Lambda$CDM is the best model, while we have definite evidence against both PADE and CPL parameterizations.

\begin{table*}
	\centering
	\caption{The statistical results for $\Lambda$CDM model obtained in different steps of analysis using different training samples. Note that we calculate the values of $\chi^2_{t}$ using test sample in each of the steps. }
	\begin{tabular}{c  c  c  c  c  c  c c c c c c}
		\hline \hline
	Sample	& Parameter  & step-1 & step-2 & step-3 & step-4 & step-5 & step-6 & step-7 & step-8 & step-9 & step-10 \\
		\hline
		& $\Omega_{\rm m0}$ & $0.2828$  & $0.2795$  & $0.2881$  & $0.2761$  & $0.2793$ & $0.2822$  & $0.2825$ & $0.2767$  & $0.2775$ & $0.2754$   \\
		\\
		Training & h & $0.6892$  & $0.6907$  & $0.6865$  & $0.6890$ & $0.6883$  & $0.6844$  & $0.6856$  & $0.6900$  & $0.6897$  & $0.6944$   \\
		\\
		& $\chi^2_{min}$ & $967.98$  & $938.72$  & $926.11$  & $940.11$  & $949.38$ & $926.59$ & $974.65$ & $935.23$  & $936.82$  & $935.16$  \\
		\hline 
	   Test & $\chi^2_{t}$ & $93.30$ & $121.42$  & $134.33$  & $119.72$  & $111.92$  & $133.20$  & $86.22$  & $124.03$  & $122.41$  & $124.48$  \\
	   \\
	      & $\chi^2_{tot}$& & & & & $1171.02$\\

		\hline \hline
	\end{tabular}\label{tab:cvl}
\end{table*}

\begin{table*}
	\centering
	\caption{The statistical results for CPL parameterization obtained in different steps of analysis using different training samples. Note that we calculate the values of $\chi^2_{t}$ using test sample in each of the steps.. }
	\begin{tabular}{c  c  c  c  c  c  c c c c c c}
		\hline \hline
	Sample	& Parameter  & step-1 & step-2 & step-3 & step-4 & step-5 & step-6 & step-7 & step-8 & step-9 & step-10 \\
		\hline
		& $\Omega_{\rm m0}$ & $0.1689$ & $0.2103$ & $0.2426$ & $0.2409$ & $0.2221$ & $0.2056$ & $0.2350$ & $0.2172$ & $0.1958$ &  $0.2433$ \\
		\\
		 & h & $0.6905$ & $0.6902$ & $0.6884$  & $0.6909$  & $0.6924$ & $0.6859$  & $0.6815$ & $0.6854$ & $0.6922$ & $0.6930$   \\
		\\
 Training & $w_0$ &  $-0.9407$ & $-0.9969$ & $-1.0342$ & $-1.0057$ & $-1.0163$ & $-0.9830$ & $-1.0001$ & $-0.9768$ & $-0.9754$ & $-0.9903$ \\
		\\
		& $w_1$ & $0.9790$ & $0.8584$ & $0.7762$ & $0.5014$ & $0.8173$ & $0.8991$ & $0.7205$ & $0.7695$ & $0.8879$ & $0.4623$  \\
		\\
		& $\chi^2_{min}$ & $964.77$  & $934.96$  & $923.05$ & $938.12$ & $945.11$ & $969.59$ & $972.73$  & $955.59$ & $956.77$  & $958.69$   \\
		\hline 
	   Test & $\chi^2_{t}$ & $92.17$  & $121.62$ & $133.39$  & $119.36$ & $112.44$ & $87.32$ & $86.45$ & $102.18$ & $100.63$  & $99.96$   \\
        \\
   & $\chi^2_{tot}$& & & & & $1055.53$\\
		\hline \hline
	\end{tabular}\label{tab:cvc}
\end{table*}

\begin{table*}
	\centering
	\caption{The statistical results for PADE parameterization obtained in different steps of analysis using different training samples. Note that we calculate the values of $\chi^2_{t}$ using test sample in each of the steps. }
	\begin{tabular}{c  c  c  c  c  c  c c c c c c}
		\hline \hline
	Sample	& Parameter  & step-1 & step-2 & step-3 & step-4 & step-5 & step-6 & step-7 & step-8 & step-9 & step-10 \\
		\hline
		& $\Omega_{\rm m0}$ & $0.2526$ & $0.2226$ & $0.2683$ & $0.2369$ & $0.2210$ & $0.1989$ & $0.3137$ & $0.1971$ & $0.2071$ & $0.2226$   \\
		\\
		 & h & $0.6898$ & $0.6916$  & $0.6845$  & $0.6919$  & $0.6885$  & $0.6885$  & $0.6866$  & $0.6881$  & $0.6872$  & $0.6963$   \\
		\\
 Training & $w_0$ & $-1.0090$ & $-1.0477$ & $-1.0572$ & $-1.0638$ & $-1.0007$ & $-0.9874$ & $-0.9816$ & $-0.9392$ & $-1.0007$ & $-0.9610$   \\
		\\
		& $w_1$ & $0.2056$  & $0.4943$ & $0.7654$ & $0.1528$ & $0.8489$  & $0.7496$ & $0.8271$  & $0.8420$ & $0.5817$  & $0.6518$  \\
		\\
		& $w_2$ & $0.3678$ & $0.5860$ & $-0.1719$ & $0.8548$ & $-0.0962$ & $0.3036$ & $0.0330$ & $-0.1144$ & $0.5883$ & $-0.1377$  \\
		\\
		& $\chi^2_{min}$ & $965.81$ & $934.35$ & $923.66$ & $937.02$ & $945.41$ & $969.12$  & $970.25$ & $955.70$  & $956.27$ & $958.39$  \\
		\hline 
	   Test & $\chi^2_{t}$ & $92.51$  & $121.82$  & $134.49$  & $118.40$& $112.23$ & $87.24$ & $96.08$ & $102.36$ & $100.56$  & $100.80$  \\
        \\
   & $\chi^2_{tot}$& & & & & $1066.50$\\
		\hline \hline
	\end{tabular}\label{tab:cvp}
\end{table*}

\begin{figure*} 
	\centering
	\includegraphics[width=8cm]{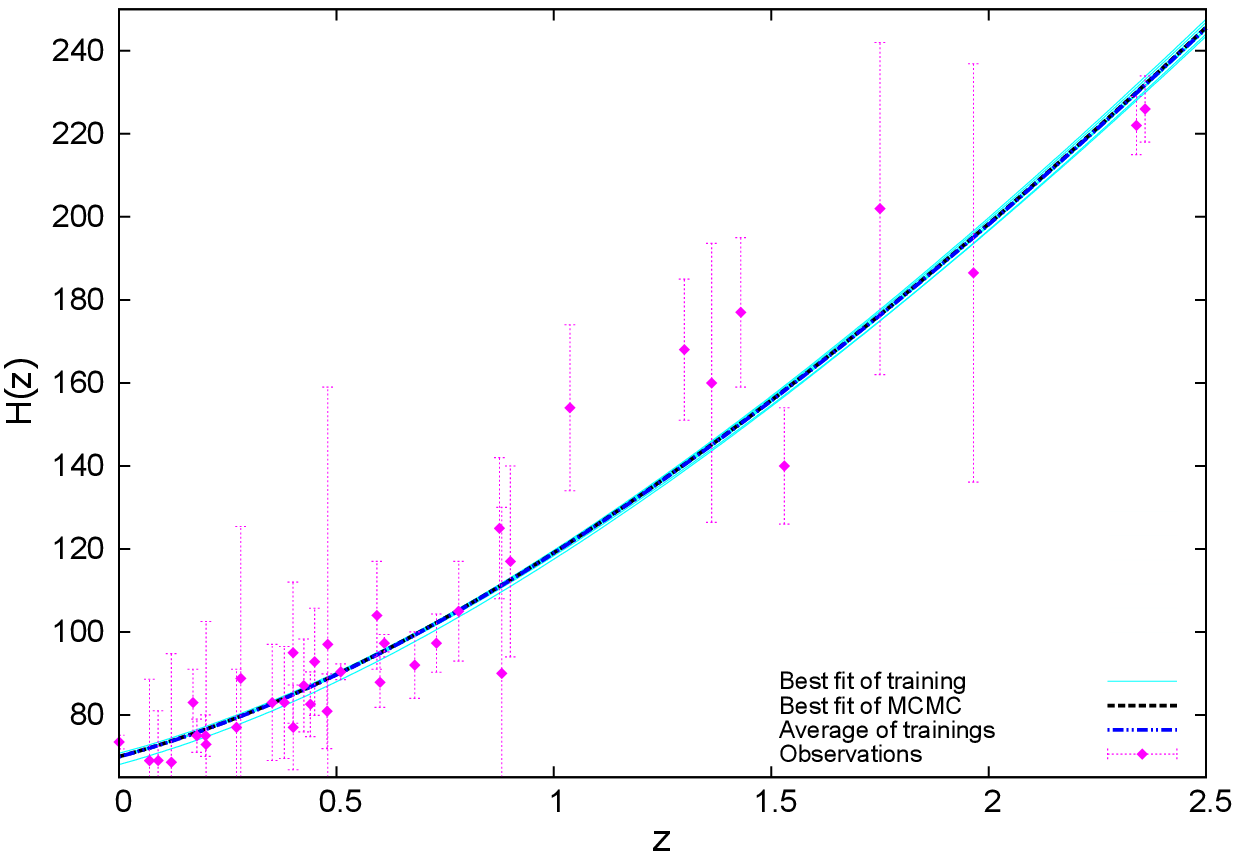}
	\includegraphics[width=8cm]{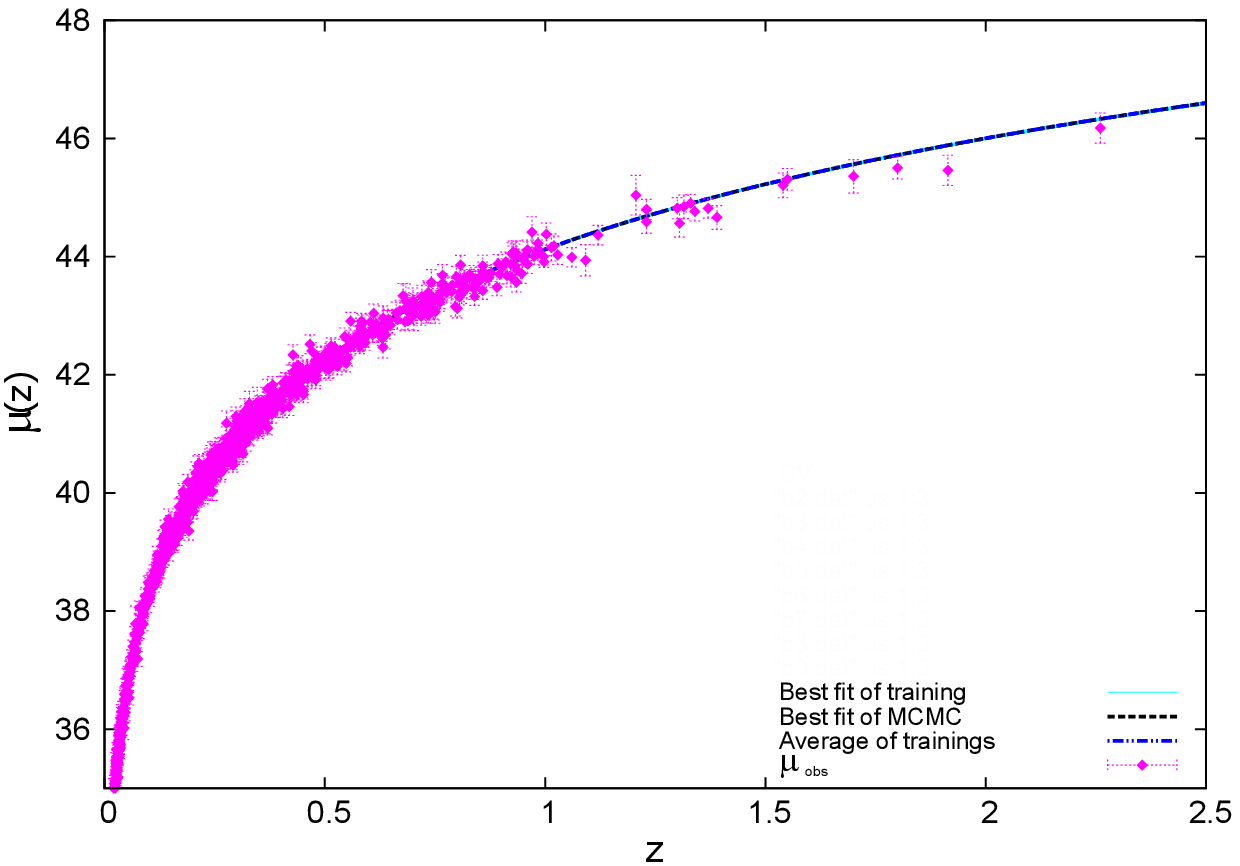}
	 \includegraphics[width=8cm]{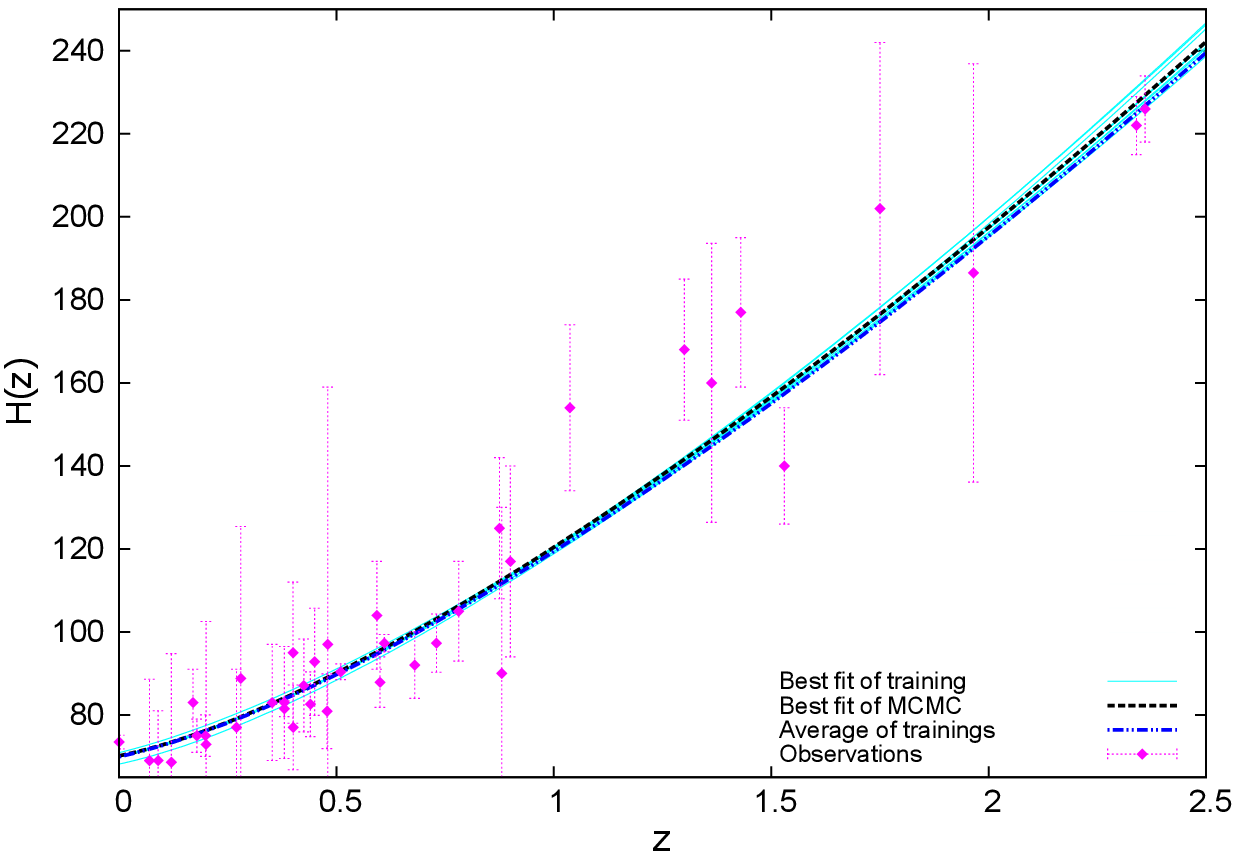}
	 \includegraphics[width=8cm]{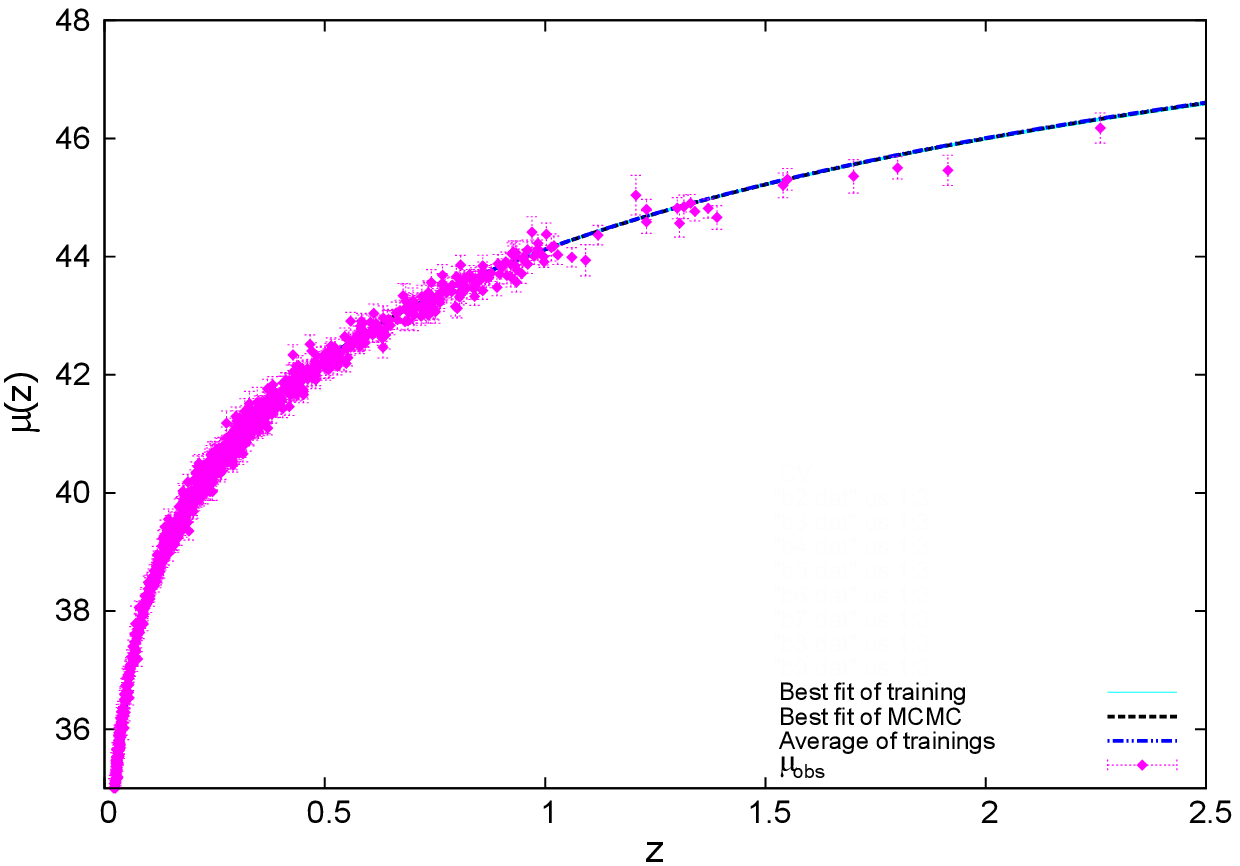}
	 \includegraphics[width=8cm]{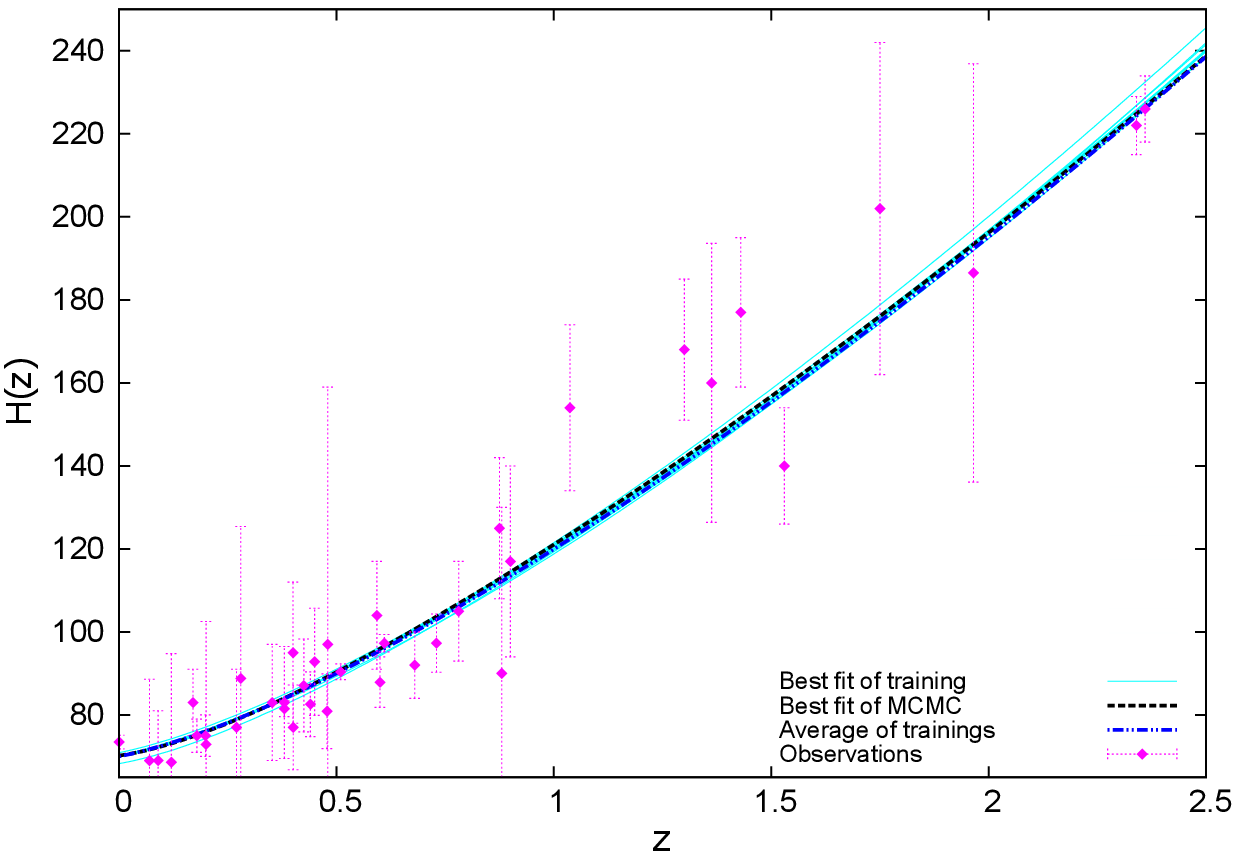}
	 \includegraphics[width=8cm]{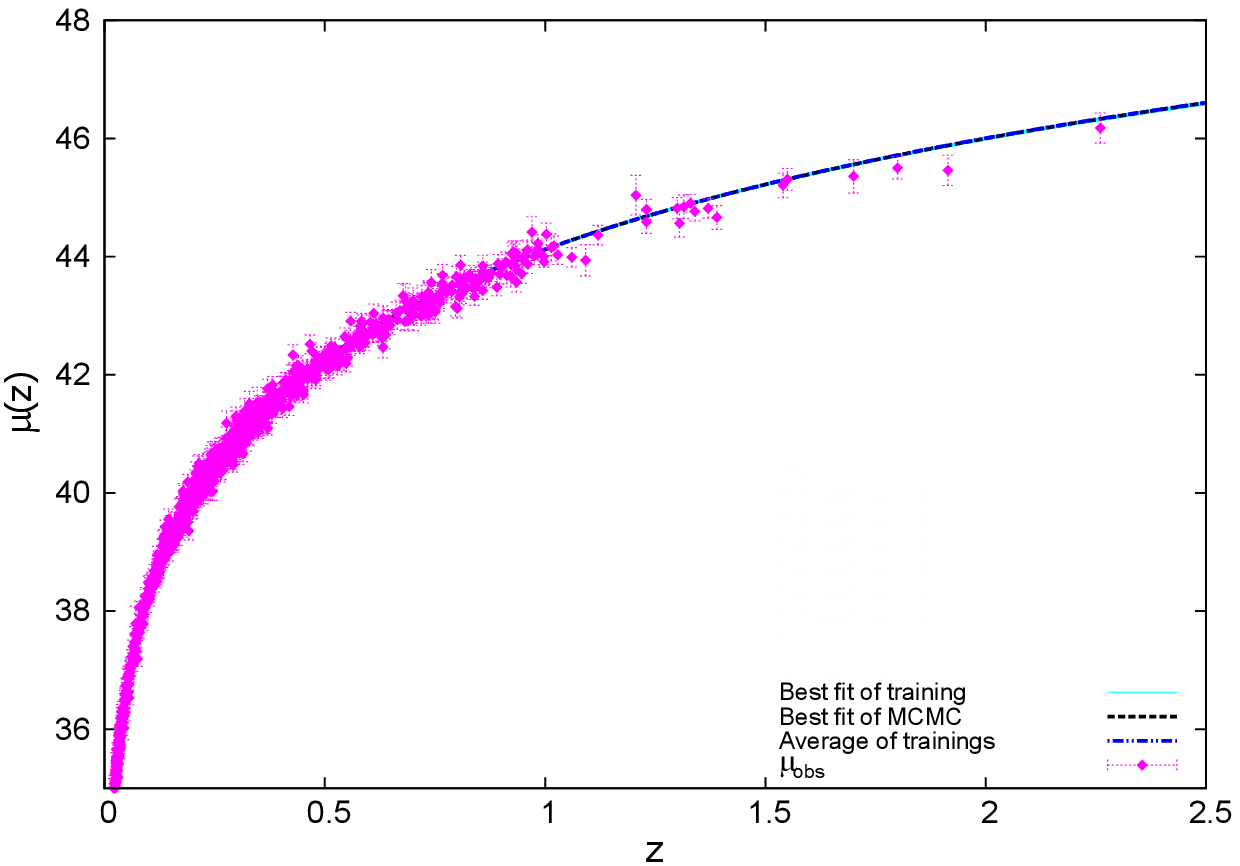}

	\caption{ Top left (right) panel: evolution of Hubble parameter (distance modulus) for $\Lambda$CDM model. Middle left (right) panel: evolution of Hubble parameter (distance modulus) for CPL parametrization. Bottom left (right) panel: evolution of Hubble parameter (distance modulus) for PADE parameterization. The curves plotted for best fit results of free parameters obtained in different analysis. The observational data points also plotted for comparison.}
	\label{fig1}
\end{figure*}

\begin{figure*} 
	\centering
	\includegraphics[width=8cm]{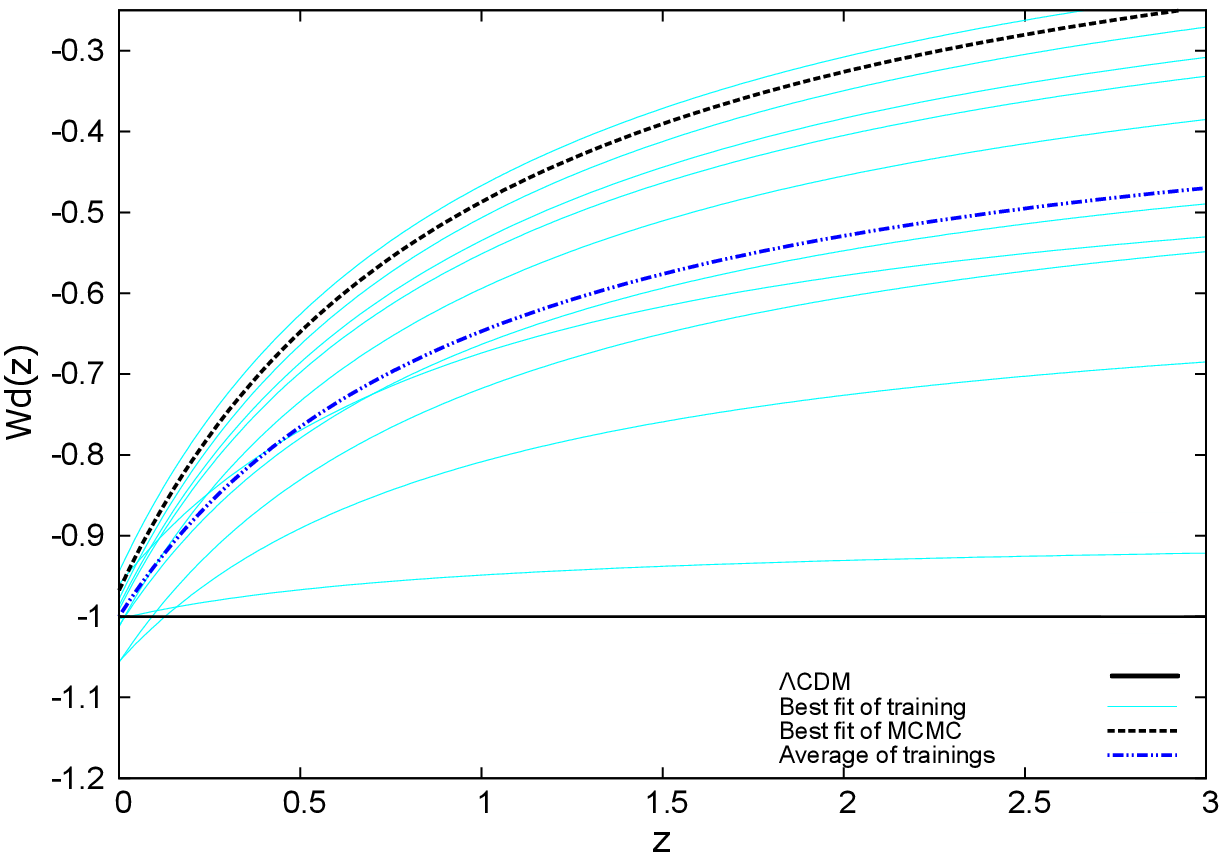}
	\includegraphics[width=8cm]{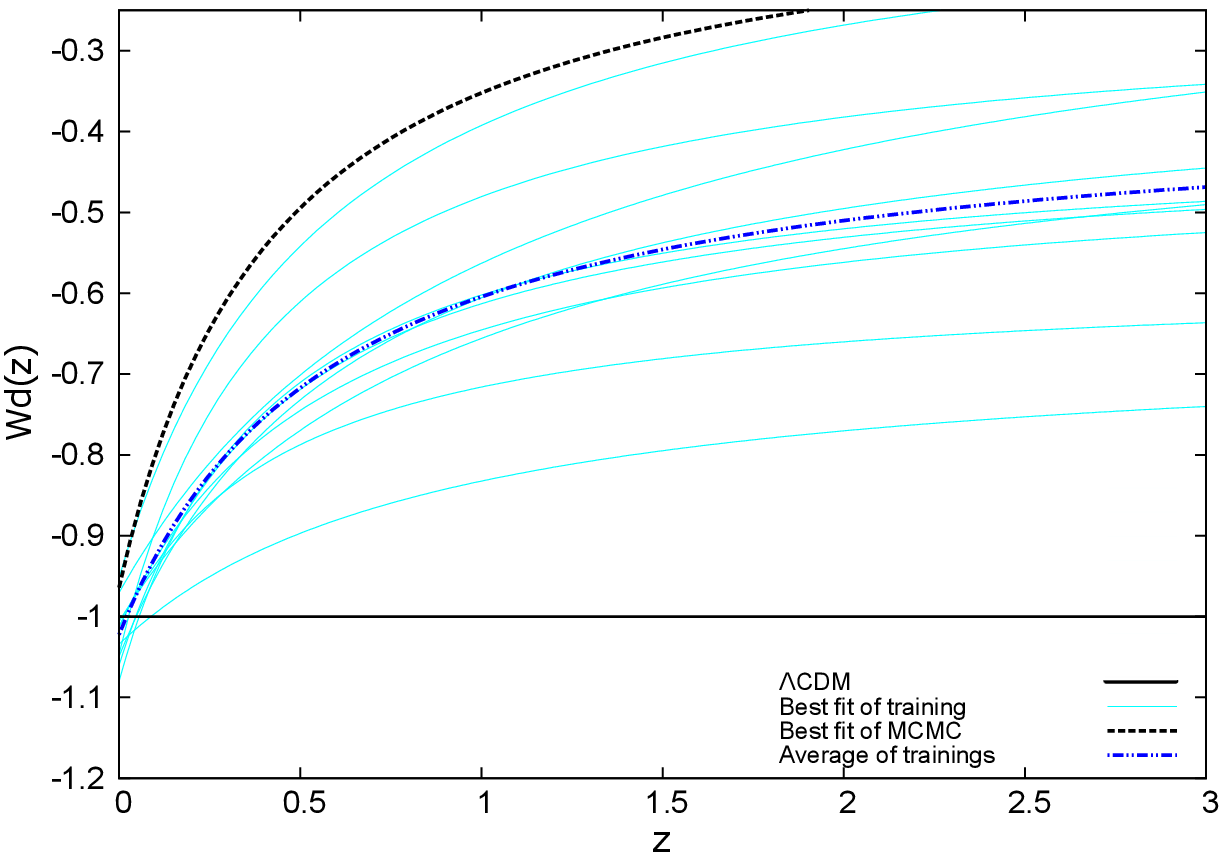}

	\caption{ Evolution of $w_d$ for different best fit parameters obtained in different analysis for CPL (PADE) parameterization plotted in the left (right) panel.}
	\label{figw}
\end{figure*}

 Now, we focus on the results of cross-validation method. As we mentioned before, our analysis was done in ten steps. In each of these steps, we perform a MCMC analysis using training sample (includes 983 data points from pantheon, H(z) and BAO sample which were selected randomly) and find the best fit parameters for DE models. Then, by choosing free parameters equal to best fits, we compute $\chi^2_{t}$ using test sample (includes remained 109 data points of pantheon, H(z) and BAO). In Tabs.(\ref{tab:cvl}-\ref{tab:cvp}) we show the results of above steps for $\Lambda$CDM model, CPL and PADE parameterizations respectively. As one can see in Tab\ref{tab:cvl} for $\Lambda$CDM sum of different $\chi^2_{t}$ values is $\chi^2_{tot}=1171.02$. We note that all of the $\chi^2_{t}$ values obtained in each step are independent from observations in test sample. We also compute the mean value of free parameters were obtained in each of steps. These mean values are $\bar{\Omega}_{\rm m0}=0.2800$ and $\bar{h}= 0.6888$ for $\Lambda$CDM model. These values are in full agreement with those we obtained from MCMC analysis using complete data sets. 
In the case of CPL parameterization, using ten different test samples we obtain $\chi^2_{tot}=1055.53$ or $\Delta\chi^2_{tot}=-115.49$ respect to $\Lambda$CDM, which means CPL is more consistent with data compared to standard model of cosmology. In this case the mean value of free parameters are $\bar{\Omega}_{\rm m0}=0.2182, \bar{h}=0.6890, \bar{w}_0=-0.9919$ and $\bar{w}_1=0.7671$.
 As the last model we compute the main values related to PADE parameterization. In this case we obtain $\chi^2_{tot}=1066.50$ and also $\bar{\Omega}_{\rm m0}=0.2291, \bar{h}=0.6893, \bar{w}_0=-1.0048 , \bar{w}_1=0.6119$ and finally $\bar{w}_2=0.2213$. Compared to other models, for PADE parameterization we have the minimum value of $\chi^2_{tot}$. In the other meaning, from cross-validation point of view, CPL parameterization with $\Delta\chi^2_{tot}=-115.49$ respect to $\Lambda$CDM is the best model and PADE parameterization with $\Delta\chi^2_{tot}=-104.52$ respect to $\Lambda$CDM occupies the second position in the model ranking. $\Lambda$CDM as the worst model, with the biggest value of $\chi^2_{tot}$ placed in the bottom position of the ranking. We plot the evolution of Hubble parameter $H(z)$ for each of the models under study in the left panels of Fig.\ref{fig1}. In each of the panels we plot different curves using the best fit parameters obtained by training samples (Tabs.\ref{tab:cvl},\ref{tab:cvc} and \ref{tab:cvp}), the mean values of free parameters and finally those we obtained in MCMC analysis. In similar fashion, we have plotted the evolution of theoretical distance modulus $\mu_{the}(z)$ for different models under study in the right panels of Fig.\ref{fig1}. Here we have  $\mu_{the}(z)=5\log [(1+z)\int^z_0 \frac {dx}{E(x)}]-5\log h +42.384$.\\
 As one can see in $H(z)$ plots (left panels) the evolution of $H$ for best fit parameters of MCMC and the mean value parameters obtained from training samples behave similarly. These two curves move between those we plotted upon the results of different training samples. This situation also exist for different curves of $\mu_{the}(z)$ which means that all of the models with their best fit parameters have same behaviors. As a main cosmological parameter, we plot the evolution of EoS parameter of DE models under study in Fig.\ref{figw}. In the left (right) panel we plot $w_d$ for different conditions of CPL (PADE) parameterization. In the both of panels we also plot $w_d=-1$ as the EoS of $\Lambda$CDM for comparison. In both of the panels we observe that the results of MCMC analysis lead to grater value of $w_d$ compared with those of mean values of parameters. Nevertheless, $w_d$ in all of conditions of CPL and PADE parameterizations behave in same manner. In all conditions, $w_d(z=0)$ is very close to that of cosmological constant ($w_d=-1$) and in higher redshifts $w_d$ decouples from phantom line and increases by $z$.\\

\section{Conclusion}\label{sect:conc}
In this work we investigate different scenarios of DE, the cosmological constant as a main candidate for playing the role of DE and two DE parameterizations as alternatives for cosmological constant. We use the CPL and PADE parameterizations for describing the form EoS parameter of DE. In order to compare these different cosmologies and select the best model, there are many different approaches which have been used in literature. As the main goal in this paper we select some of these well known approaches, the AIC, BIC, DIC criteria and the Bayesian evidence, and compare them with another approach for model selection named cross-validation. Among different methods which there are for cross-validation, we use $k$-fold cross-validation method by setting $k=10$. In this way we apply three sets of latest observational data points, the Pantheon sample as the latest 1048 type Ia supernovae observations from \citep{Scolnic:2017caz} and the latest measurements from cosmic chronometers for $H(z)$ from \citep{Farooq:2016zwm} and the BAO sample.\\
Firstly, in order to compute the values of AIC, BIC and DIC, we implement a joint likelihood analysis using all of the data points. As the result, we observe that PADE parametrization with $\chi^2_{min}=1056.6$ has the minimum value of $\chi^2$ and $\Lambda$CDM with $\chi^2_{min}=1061.4$ has the greatest value of $\chi^2$. Using these values, we compute the related AIC and BIC for different DE models under study. Based on AIC results we observe that $\Lambda$CDM cosmology is the best model, but the performance of the other models are also very good. Upon AIC, we found that there are
\textit{Significant support} to CPL and PADE models. Assuming the computed BIC values, we saw that $\Lambda$CDM with $BIC=1075.4$ is the best model and PADE parametrization with $BIC=1091.6$ is the worst one. Upon these values we found \textit{Very Strong evidence} against both CPL and PADE parametrization. Although, AIC results indicate that all of the models under study are consistent with applied data samples, but from the BIC point of view we found huge differences among the performance of the models. Calculating the value of DIC for different models under study, we observe a big conflict in results of various criteria.
In this case, PADE with $DIC=1065.9$ is the best model, while $\Lambda$CDM with $DIC=1078.3$ is the worst model. The value of $\Delta DIC=12.4$ indicate that there is a \textit{Strong evidence} against $\Lambda$CDM. Based on the Bayesian inference analysis, our results indicate that $\Lambda$CDM is the best model, while for CPL (PADE) parameterization we have $\Delta \ln \epsilon= 2.7$ ($\Delta \ln \epsilon= 2.2$) which means there is \textit{Definite evidence} against these models.\\
Finally, we do our new analysis, the cross-validation to compare the models. As one can see in Tabs.(\ref{tab:cvl}-\ref{tab:cvp}), cross-validation results indicate that CPL cosmology with $\chi^2_{tot}=1055.03$ has the minimum value of $\chi^2_{tot}$ among investigated models and thus, is the best model. On the other hand $\Lambda$CDM model with $\chi^2_{tot}=1171.02$ has the worst performances among different models. Although, we found conflict between the results of AIC and BIC and those of DIC, but now we find a greater conflict between Bayesian evidence results and those of cross-validation. Here, the main question is " which of these methods for model selection are the best?" or equivalently, "which of these cosmologies are the best one?".
From cosmological point of view we know that there are different theoretical problems against cosmological constant which dynamical DE can solve or alleviate them. Furthermore, as we mentioned in Sec.\ref{sect:int} $\Lambda$ cosmology plagued with different observational tensions in estimation of some of main cosmological parameters. It was observed in some studies that dynamical DE scenarios can reduce or even solve some of these tensions \citep[see also][]{Rezaei:2019xwo,Pan:2019hac}. On the other hand, from model selection point of view, we show that AIC just can help us to identify the most useful model for prediction, while BIC provides good results if in the set of the models under consideration is one favored model. Furthermore, BIC obtained using the values of $\chi^2_{min}$ which calculated upon the value of free parameters that constrained with observational data. It is important to note that the data samples which have used to constrain free parameter are exactly same as those have used to compute $\chi^2_{min}$. This point can affect the results which we obtain using BIC. In contrast with AIC and BIC criteria, instead of using just the best fit likelihood, DIC uses the whole sample. Furthermore, AIC and BIC penalize all the involved parameters, while DIC penalizes just those parameters which contribute to the fit in an actual way.
On the other hand, Bayesian inference not only determines the free parameters of the model, but also provides a direct way to compare different models. Moreover, the Bayesian evidence for model selection has been widely used in cosmology.\\
As the last method of model selection we have used cross-validation. In cross-validation method, the data sample which we use for computing $\chi^2$ value is independent from those we use for training the model. This item can significantly improve the position of cross-validation against other approaches, especially BIC and AIC criteria. Combining all of these and assuming the numerical results we obtained in this work, we can say that choosing cross-validation and DIC methods for model selection leads to precedence of dynamical DE scenario, while AIC, BIC and Bayesian evidence lead to precedence of $\Lambda$ cosmology. Therefore, we conclude that using cross-validation and DIC methods for model selection can lead to different results compared to BIC and AIC. Our results, can be examined using other different data samples or by applying cross-validation for comparing other different cosmological models. Cross-validation besides other methods, can be considered as a potential candidate for model selection in cosmology in further investigations.

\section{Acknowledgements}
The authors gratefully thank to the Referee for the constructive comments and recommendations which definitely help to improve the readability and quality of the paper.

\bibliographystyle{aasjournal}
\bibliography{ref}
\label{lastpage}

\end{document}